# AN ADAPTIVE EMBEDDED ARCHITECTURE FOR REAL-TIME PARTICLE IMAGE VELOCIMETRY ALGORITHMS.

*Alain Aubert, Nathalie Bochard, Virginie Fresse*

Laboratoire de Traitement du Signal et Instrumentation
CNRS UMR 5516, bat F. 18 rue Benoit Lauras, 42000 Saint Etienne, France.
phone: + (0033) 477 91 57 94, fax: + (0033) 477 91 ,
email: {alain.aubert, nathalie.bochard, virginie.fresse}@univ-st-etienne.fr

## ABSTRACT

Particle Image Velocimetry (PIV) is a method of imaging and analysing fields of flows. The PIV techniques compute and display all the motion vectors of the field in a resulting image. Speeds more than thousand vectors per second can be required, each speed being environment-dependent. Essence of this work is to propose an adaptive FPGA-based system for real-time PIV algorithms. The proposed structure is generic so that this unique structure can be re-used for any PIV applications that uses the cross-correlation technique. The major structure remains unchanged, adaptations only concern the number of processing operations. The required speed (corresponding to the number of vector per second) is obtained thanks to a parallel processing strategy. The image processing designer duplicates the processing modules to distribute the operations. The result is a FPGA-based architecture, which is easily adapted to algorithm specifications without any hardware requirement. The design flow is fast and reliable.

## 1. INTRODUCTION

Particle Image Velocimetry PIV is a method of imaging and analysing fields of flow in critical environment. The initial groundwork for a PIV theory was laid down by Adrian [1] who described the expectation value of the auto-correlation function for a double exposure continuous PIV image. Nowadays many techniques exist and they all remain computing intensive [2]. Traditional systems are therefore not suitable for real-time PIV applications as they cannot achieve the required high performance and be integrated in the critical environment. From an algorithmic point of view, several parameters depend on the experimental environment. The size of images, camera frequency and other information are tailored for a given environment as they depend on the characteristics (size and speed) of the fluid.

As a result, an embedded dedicated architecture must be designed for real-time PIV algorithms. This system must be adapted to the given algorithm specifications to meet the constraints without requiring a complete system redesign when the critical environment changes. To date, FPGAs are increasingly used in embedded systems as they can achieve high-performance in a small footprint. As modern FPGA integrate many different heterogeneous resources on one single chip, the complete image processing algorithm can be implemented without any other external resources. More importantly, the reconfigurable aspects of FPGA give the circuit the versatility to change its functionality according to the algorithm requirements.

Essence of this work is to propose an adaptive embedded architecture dedicated to real-time PIV algorithms. According to the given constraints and the required results, the designer changes the FPGA-based architecture with only few modifications and without any hardware requirement.

This paper is organised into 4 further sections. Section 2 presents the Particle Image Velocimetry algorithm. Section 3 introduces the proposed FPGA-based system with the global structure. Experimental results are presented in section 4 and section 5 concludes the paper.

## 2. PIV ALGORITHM

Particle Image Velocimetry is a technique for flow visualisation and measurement. The fluid motion is measured thanks to particles seeded in the flow.

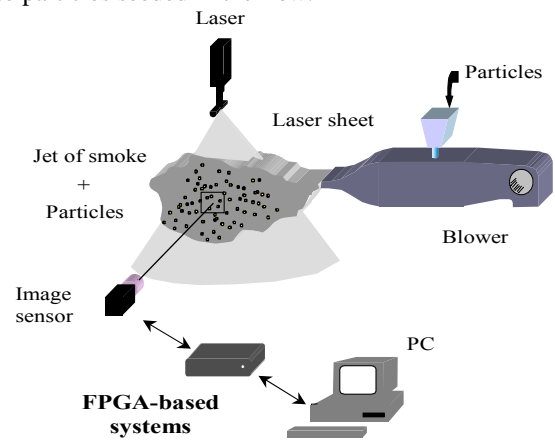

**FIGURE 1. A real-time PIV system.**

Indeed, small particles are used as markers for motion visualisation in the studied flow as shown in figure 1. This is a non invasive measure as the sizes of particles do not alter the flow or fluid properties [3].

Two images are taken from one or two cameras within a short time interval t and t+Δt. In our application two single exposure image frames are recorded by one camera only. Images recorded by the camera are divided into small sub-regions called interrogation areas or interrogation windows.

From the interrogation window of the second image is extracted a pattern. This pattern is shifted in the corresponding interrogation window in image1 and both are direct cross-correlated as shown in equation 1.

$$F(i,j) = \sum_x \sum_y s1(x,y) \times s2(x-i, y-j) \quad (1)$$

where s1 and s2 respectively represent the grey levels of the interrogation windows from images 1 and 2.

The resulting output on the correlation plane is a single peaked function where the peak represents the displacement of the particles (figure 2). Direction of the displacement is determined unambiguously because the images from exposure 1 and 2 are recorded separately.

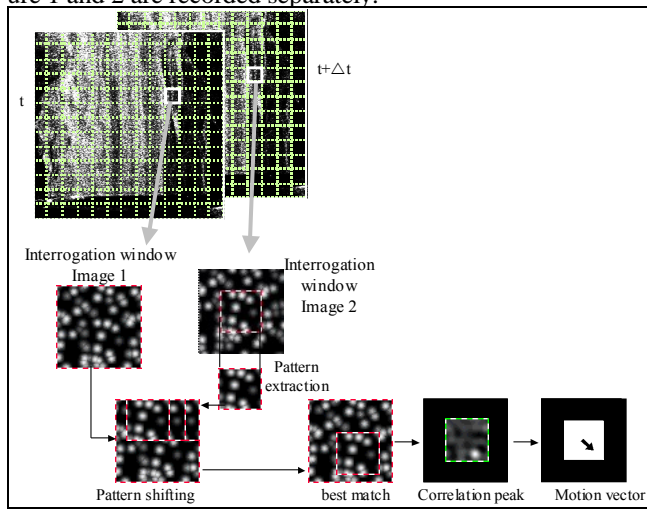

**FIGURE 2. 2 single exposure input sub-regions and the corresponding output cross-correlation plane. The location of the single bright correlation peak from the origin is the average displacement.**

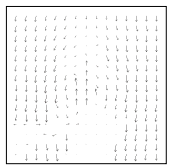

**FIGURE 3. Resulting image of PIV**

The resulting image is a set of vectors indicating the flow displacement as shown in figure 3.

Traditional technique using grey-level images is adapted to binary direct cross-correlation to ensure an easier implementation on programmable logical circuits. A binary direct cross-correlation gives efficient results if preceded by a suitable binary operation as shown in [4]. Multiplications used in equation (1) are replaced by XNOR logical operations:

$$F(i,j) = \sum_x \sum_y s1(x,y).XNOR.s2(x-i, y-j) \quad (2)$$

PIV algorithm is suitable for parallel processing as the direct cross-correlation computation is highly parallelisable. Two cross-correlated interrogation windows are independent of each other. A unique operation is computed simultaneously on different interrogation windows. These complex computations are therefore good candidates for a hardware real time implementation.

First characteristic of PIV applications is unbalanced data flows between input and output as showed in figure 4. The input data flow captures several images meaning that input data correspond to thousand of pixels. The output data flow represents a little number of vectors. Another characteristic is the data parallelism inherent to this application. The correlation is duplicated several times (according the sizes of image and interrogation window) and processing operates on local pixels.

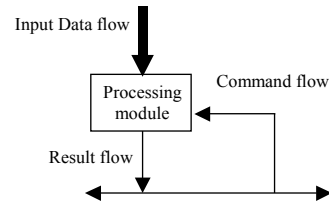

**FIGURE 4. Main characteristics of PIV applications**

Using these characteristics, a generic structure for PIV algorithms is defined.

## 3. DEDICATED ARCHITECTURE

The FPGA-based system presents a Globally Asynchronous Locally Synchronous (GALS) architecture (figure 5). For all GALS structures, each module runs at its own frequency and communicate asynchronously with a handshake protocol. All synchronous blocs units are therefore independent. As a result, a modification required for a specific module does not alter the rest of the architecture. For example, a CMOS image sensor must be replaced by a camera with a higher frequency. The modifications only concern the acquisition module as other modules are independent.

A main characteristic is the communication ring for control data and output data. All modules are inserted around this ring by means of an asynchronous wrapper structure. This wrapper is a 4-phase handshake protocol and is identical for all types of module. The control remains identical and trusty operations are insured. The advantage of this ring includes minimizing latency as several commands can run simultaneously inside the ring. Empty commands can also be sent by the control module. This empty command can be used by any module to send some results back to the control module. As all modules are inserted around the communication ring, there are no limits in the number of inserted modules. And a module insertion does not modify the global structure. The advantage is a unique communication protocol whatever the number and type of modules.

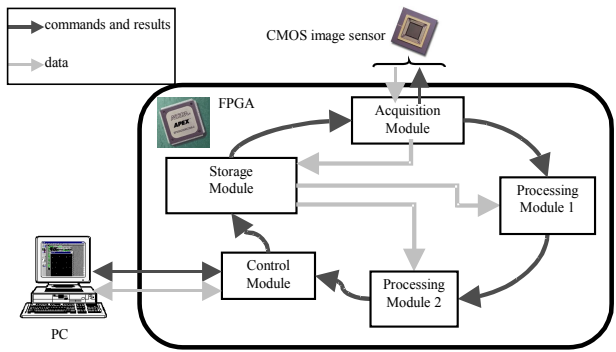

**FIGURE 5. Adaptive FPGA-based architecture for real-time PIV applications.**

This structure handles different type of operations which are required in PIV applications. All these operations are designed in specific modules [5].

### 3.1 Different type of modules

- The **Control Module** sends commands to each module through the communication ring. These commands activate predefined macro functions in the target module. The integrity and the acceptation of the commands are checked with a flag inserted in the same command that returns in the control module. As all commands are sent from this module, the scheduling is specified in the control module. In the same way, this module receives resulting data transferred coming from processing modules.

- The **Acquisition Module** produces all CMOS image sensor commands and receives CMOS image sensor data. One part takes the 10 bit-pixel data from the sensor and generates one bit data thanks to a binarisation operation. As the binarisation depends on the critical environment (i.e. the position of the camera and the acquired scene), this pre-processing operation can use either a unique threshold for the entire image or a local and adaptive threshold for each sub-region.

32 binary data are packed and sent to the storage module. All commands dedicated to acquisition operations (windowing, configuration, pre-processing…) are received from the Control Module through the ring.

- The **Storage Module** stores incoming images from the Acquisition Module. Memory banks are FPGA-embedded memories, and writing and reading cycles are supervised by the control module.

- The **Processing Module** contains the logic, which is required for the direct cross-correlation operations. The result from the image processing is then sent to the control module by means of the communication ring. More than one processing module can be used for the parallel direct cross-correlation operations.

In the GALS structure, each type of modules runs at their own frequency. For the following implementations, the used frequency are given in table I.

TABLE I: Frequencies per type of module.

| Modules | Processing | Acquisition | Storage | Control |
|---|---|---|---|---|
| Frequency (MHz) | 100 | 10 | 100 | 150 |

### 3.2 Asynchronous wrapper

An asynchronous wrapper, figure 6 is designed to accept multi-clock domains. Each module is a synchronous module running at its own frequency. Communications between modules are asynchronous and they use a single-rail data path 4-phase handshake. The wrapper includes two independent asynchronous units. One receives frames from the previous module and the other one sends frames to the following module at the same time.

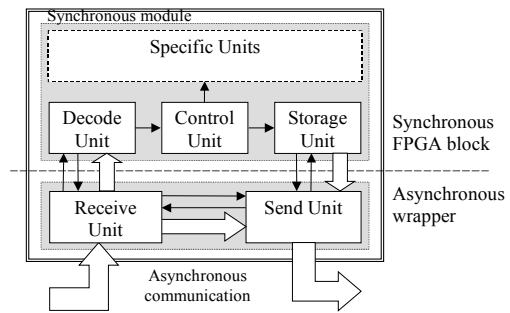

**FIGURE 6. The asynchronous wrapper structure.**

### 4. IMPLEMENTATIONS AND RESULTS

For the following implementations, the size of images is 320*256 with 80 interrogation windows (with a size of 32*32 pixels). The FPGA-based system is a NIOS II board with a Stratix II 2S60 FPGA. Image data are acquired in a 100ns CCD sensor.

The initial architecture contains 4 modules, one module for each type of operation. From this structure is added one or more processing modules in the communication ring that makes the execution time faster. The binary cross-correlations operations are then proceed in a parallel design strategy. As a consequence all processing module execute a binary cross-correlation operation several times. With an architecture that contains one processing module, this module performs all cross-correlation operations, and generates 80 vectors. With an architecture that contains two processing modules, both modules execute half of the cross-correlation oper-

ations, each processing module giving 40 vectors. And so on…

The proposed architecture is implemented with one processing modules up to 6 processing modules. The implementation results are given in table II.

TABLE II : Implementation results.

| Nb of processing modules | Image/sec. | Pixel clock frequency (MHz) | Nb of vectors per second |
|---|---|---|---|
| 1 | 204 | 16.8 | 16 393 |
| 2 | 403 | 33.0 | 32 258 |
| 3 | 571 | 46.8 | 45 714 |
| 4 | 757 | 62.1 | 60 606 |
| 5 | 925 | 75.9 | 74 074 |
| 6 | 1 063 | 87.1 | 85 106 |

Observation is as follow: the binary cross-correlation operations are distributed onto all processing modules. The parallel design strategy equally distributes the operations for each module. As a result, the number of vectors per second increases when the number of processing modules increases. This FPGA-based system can handle high-speed constraints such as more than 80000 vectors/second and be adapted to high-speed applications. With 6 processing modules, this system can work with a speed of 1 000 images/second (for a 320*256 image) and more than 330 images per second for a 512*512 image.

As the processing modules are inserted around the communication ring, there are no restrictions in the number of processing modules. The communication remains unchanged and resources required for the communication between modules are few.

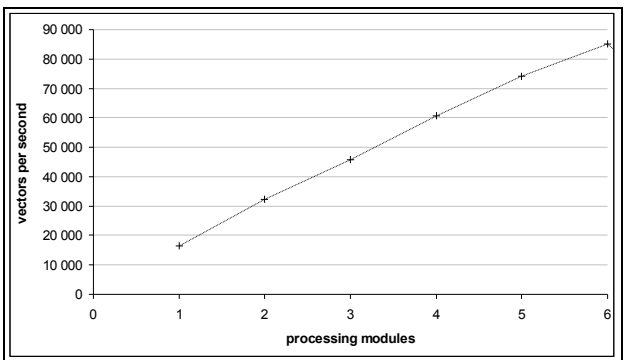

**FIGURE 7. Number of vector per second per number of processing modules.**

Therefore, this FPGA-based system can integrate a high number of processing modules. As shown in figure 7, the number of processing depends on the specified speed (number of vectors per second). The image processing designer numbers how many processing modules are required and inserts them in the architecture.

Theoretically, the number of processing modules is unlimited. The number of required resources depends on the number on processing modules. Therefore this number of processing modules will depend on the available FPGA resources. On the other hand, the communication ring will slow down the system for a very high number of processing modules. Indeed, the execution time decreases for an increasing number of modules but the communication time between neighboured modules remains unchanged. Present work consist in evaluating the limit (i.e. the number of processing modules) for which the communication ring will slow down the system.

## 5. CONCLUSION AND PERSPECTIVES

As a result, this FPGA-based system presents a generic architecture that can be easily adapted to any real-time PIV algorithms, which uses the cross-correlation technique. All functional blocks remain unchanged only the number of processing modules changes.

According to the speed, the image processing designer evaluates the number of processing modules. Then, he duplicates an identical processing modules several times around the communication ring without any functional modifications. All other type of blocks remain unchanged that makes the design flow fast and reliable.

Future work consists in demonstrating that changing any external devices does not alter the rest of the structure. The CMOS sensor will be therefore replaced by another external devices such as a camera or a CCD sensor with another frequency. The GALS structure ensures to integrate any modules running at a different frequency. In this paper, only the speed changes. Perspectives are to prove that this generic architecture can be also easily modified when the size of the image changes or the interrogation window.